\documentclass{article}
\usepackage{xcolor,soul}
\usepackage{hyperref}
\usepackage{graphicx}
\graphicspath{{./figures/}}
\usepackage{footmisc}
\usepackage{float}
\usepackage{amsthm}
\usepackage{tabularx}

\usepackage{graphicx}
\graphicspath{ {figures/} }
\usepackage{subcaption}
\usepackage{bm}
\newcommand{\approxtext}[1]{\ensuremath{\stackrel{\text{#1}}{\approx}}}

\usepackage{arxiv}

\usepackage[utf8]{inputenc} % allow utf-8 input
\usepackage[T1]{fontenc}    % use 8-bit T1 fonts
\usepackage{hyperref}       % hyperlinks
\usepackage{url}            % simple URL typesetting
\usepackage{booktabs}       % professional-quality tables
\usepackage{amsfonts}       % blackboard math symbols
\usepackage{nicefrac}       % compact symbols for 1/2, etc.
\usepackage{microtype}      % microtypography
\usepackage{lipsum}		% Can be removed after putting your text content
\usepackage{graphicx}
\usepackage{doi}
\usepackage{amsmath}
\usepackage{appendix}

\title{Eliciting Informative Priors by Modelling Expert Decision Making}

\author{ \href{https://orcid.org/
0000-0002-7970-6342}{Julia R. Falconer}\thanks{Julia Falconer's research is funded through the University of Waikato Doctoral Scholarship} \\
	Department of Mathematics,\\
	University of Waikato, \\
	Hamilton, New Zealand \\
	\texttt{jrg22@students.waikato.ac.nz} \\
	\And
	{Eibe Frank} \\
	Department of Computer Science,\\
	University of Waikato, \\
	Hamilton, New Zealand\\
	 \\
	\And
	{Devon L. L. Polaschek}\\
	School of Psychology,\\
	University of Waikato, \\
	Hamilton, New Zealand
	\And
	{Chaitanya Joshi}\\
	Department of Statistics,\\
	University of Auckland, \\
	Auckland, New Zealand 
}

\renewcommand{\shorttitle}{Eliciting Informative Priors by Modelling Expert Decision Making}

\hypersetup{
pdftitle={Eliciting Informative Priors from Expert Decision Making},
pdfauthor={Julia R. Falconer, Eibe Frank, Devon L. L. Polaschek, Chaitanya Joshi},
pdfkeywords={First keyword, Second keyword, More},
}

\begin{document}
\maketitle

\begin{abstract}

This article introduces a new method for eliciting prior distributions from experts. The method models an expert decision-making process to infer a prior probability distribution for a rare event $A$. More specifically, assuming there exists a decision-making process closely related to $A$ which forms a decision $Y$, where a history of decisions have been collected. By modelling the data observed to make the historic decisions, using a Bayesian model, an analyst can infer a distribution for the parameters of the random variable $Y$. This distribution can be used to approximate the prior distribution for the parameters of the random variable for event $A$. This method is novel in the field of prior elicitation and has the potential of improving upon current methods by using real-life decision-making processes, that can carry real-life consequences, and, because it does not require an expert to have statistical knowledge. Future decision making can be improved upon using this method, as it highlights variables that are impacting the decision making process. An application for eliciting a prior distribution of recidivism, for an individual, is used to explain this method further.
\end{abstract}

% keywords can be removed

\pagebreak

\section{Introduction}
Beginning with some prior knowledge (a prior probability distribution), Bayesian inference updates the prior by taking information from observed data (a likelihood) to build a posterior distribution over the parameters of interest, $\theta$: 
	\begin{equation} \label{BayesianEquation}
	p(\theta | y) \propto p(\theta)p(y|\theta),
	\end{equation}

A prior distribution that has minimal influence on the posterior distribution, a 'non-informative' prior, is often used. Where there is large amounts of data, the choice of prior is largely irrelevant since the likelihood dominates the posterior distribution. However, if there is limited data, the influence from the likelihood becomes minimal, producing a posterior that relies heavily on the prior information. For such instances, an informative prior distribution could be used \cite{zyphur2015bayesian}. \\

\begin{table}[h]
	\caption{\label{Tabledef} Definitions expanded from a table from  \cite{falconer2021methods}}.
	\centering
	\begin{tabular}{l  p{.8\textwidth}}
 \hline
		\toprule
		%\multicolumn{2}{c}{Part}                   \\
		%\cmidrule(r){1-2}
		\textbf{Name}     & \textbf{Description}\\ \hline
		\midrule
		\emph{Prior Elicitation} & The process of obtaining knowledge from a source to form a prior distribution that can be used for further Bayesian analysis.     \\
		\hline
 \emph{Expert} & An individual (or a group of individuals) who has extensive knowledge on a certain subject matter. The expert is also referred to as the decision maker in this text.\\ \hline
 \emph{Decision Maker} & The individual who performs a decision making task. In most cases, the Decision Maker and the Expert will be the same individual.\\ 
\hline
 \emph{Analyst} & An individual who performs the task of forming a prior distribution using prior elicitation techniques. \\
 \hline
 \emph{Facilitator} & An individual who performs the task of eliciting knowledge. In some cases, the Facilitator and the Analyst may be the same individual.\\
 \hline
		\bottomrule
	\end{tabular}
	
\end{table}

We consider scenarios exhibiting an event, $A$, that is of serious consequence and where data on $A$ is limited as the event rarely occurs.  An analyst (see Table \ref{Tabledef} for definitions used throughout this paper) wishes to obtain an informative prior distribution for $A$. Although there may not be any data on $A$, there may be some other related source of information that can be used to obtain a prior for $A$. The most common way to do this is to elicit a distribution from an expert in the relevant field of interest. Methods to obtain an informative prior distribution from an expert are described in \cite{falconer2021methods}, which assigns methods to three categories; 1) Direct Interrogation Methods, 2) Indirect Interrogation Methods, and 3) Graphical/ Visual Methods. Direct Interrogation methods \cite{o2006uncertain,galway2007subjective,jenkinson2005elicitation} involve asking experts about the probability distributions directly. This can be challenging because experts must first have a firm grasp of probability theory and distributions. There are circumstances where an expert can first be taught key probability concepts \cite{o2006uncertain,thomas2020probabilistic,casement2018graphical}, but this can prove difficult and create inaccurate prior distributions \cite{kadane1998experiences,wang2013expert}.  This issue can also be seen in some graphical/visual methods \cite{falconer2021methods}. Indirect Interrogation methods have been introduced to help combat the requirement of experts needing knowledge of probability theory. Indirect Interrogation methods involve asking the expert questions that are not directly based on the probability distributions themselves, but instead are easy for the expert to comprehend. From there, an analyst will use mathematical logic to infer a prior distribution. Two examples of Indirect Interrogation that display the simplicity of questioning are: getting the expert to place bets on which event they think is more likely \cite{winkler1967quantification} and getting the expert to rank the likelihood of events \cite{eckenrode1965weighting,edwards1994smarts,wang2013expert}. As highlighted in \cite{falconer2021methods}, some Indirect Interrogation methods can be thought of as hypothetical decision-making tasks. Hypothetical decision-making implies that whether the decision is correct or incorrect has no real consequence for the expert. Therefore, prior elicited in this way may not accurately reflect the expert's thinking in real life.  \\

The use of experts during the process of elicitation has the added complexity of introducing cognitive and motivational biases. In Direct Interrogation elicitation, the simple mistake of asking a question a certain way can produce cognitive biases which influence the experts response (e.g., anchoring and adjusting \cite{kahneman1982judgment}, where values in the questions are used by an expert to anchor their response value). Prior elicitation methods that use experts may also have cognitive biases based on an expert's work experience (e.g., judgment by availability \cite{kahneman1982judgment}, where an expert will put more weight on an event just because the expert witnessed that event more recently) or, to put it more generally, an expert's life experience, that includes biases they have formed over time (e.g., gender bias, racial bias). 
Using a group of experts to elicit one prior distribution can help an analyst gain a wider view of the whole field of interest \cite{o2019expert}. A common way to do this is to get a group of experts to discuss opinions to form a consensus, however, this method can also come with cognitive biases that an analyst should be aware of, such as \textit{groupthink} \cite{janis1983groupthink}. Groupthink is where the need to reach a consensus, while maintaining harmony within the group, means individuals do not voice alternative perspectives that may be outside the social "norm" or maybe against the perspective of a strongly influential individual, skewing the group's elicited prior in one direction \cite{janis1983groupthink}. Instead of having experts reach a consensus, some methods allow analysts to combine experts' individual priors mathematically \cite{o2006uncertain} to avoid cognitive biases that are formed from group consensus, such as groupthink. Some methods can elicit a prior distribution without an expert's input by using historical data (e.g., use the posterior from a similar historical study \cite{press2009subjective}), however, in most cases this historical data will not exist. Also, historical data is not immune from the effects of biases, and it is not just an individual expert's cognitive biases that an analyst must be aware of.  Sometimes available data might encompass societal biases \cite{belenguer2022ai}. A famous example is the Correctional Offender Management Profiling for Alternative Sanction, COMPAS \cite{angwin2016machine}. COMPAS was a risk assessment tool that was used to obtain a recidivism score for defendants. Although ethnicity was not a factor in the model, the tool was still more likely to class black individuals as high risk than other individuals \cite{angwin2016machine}. This was because the model had learned from historic discriminatory court cases and enhanced the prejudices in the judicial system \cite{belenguer2022ai}. Another example is a tool that was used to rank the top five applicants based on their resumes for job vacancies at Amazon; it was found to be penalising applicants that were women and favouring those that were male \cite{dastin2018amazon,belenguer2022ai}. This was because the model learned patterns from historic data where women were not hired for positions at tech companies \cite{belenguer2022ai}. The societal biases of blacks being more likely to commit crimes and females being less adequate for specific jobs were shown in the data applied to these models and influenced the outputs. Lack of information or inadequate information can also produce a bias \cite{jargowsky2005omitted}. If the available information is heavily dominated by information on one group then it is obvious that the results produced with this information could be considered biased, like in the COMPAS example. Often the available information is tabular data, which may be missing key information that is needed to give accurate outputs. Tabular data variables may also represent multiple factors of interest that are not directly collected in the data (confounding variables), making it hard to understand what variables are truly influencing the output. 
Reducing the impact of biases on the elicited prior is a key interest in prior elicitation \cite{o2019expert,o2006uncertain}.\\

\subsection{Motivation}
We believe the key limitations of current methods are: a) the statistical knowledge required of experts to perform elicitation by Direct Interrogation methods, b) the "hypothetical" decision-making tasks in Indirect Interrogation methods that have no real-life impact and could affect the accuracy of the elicited prior, and, c) the difficulty  of identifying %or mitigating 
biases when eliciting an expert prior. We introduce a method that eliminates some of these limitations by obtaining an approximation of a prior distribution through modelling an expert's past decision-making tasks. Our method eliminates the statistical knowledge required by utilising a decision-making task that an expert performs as part of their duties. Often this decision-making task has real-life implications, meaning more importance is placed on the decision, and the experts will strive to be more accurate in their decisions.  Thus, by modelling their past decisions, we may be able to capture their thinking more accurately than methods that rely on hypothetical decision-making. Also, modelling past real-life decisions eliminates biases that could be introduced in direct interrogation methods.  Although, because we are using experts, there may still be cognitive biases affecting the elicited distribution. Modelling data from the past decision-making tasks may allow analysts to identify variables that may be considered to be inducing bias in the decision-making process. 
The method is explained further in Section \ref{sec:Methods}. We discuss ways to assess model behaviour in Section \ref{sec:PM}, with Section \ref{sec:examples} outlining a simple example application. Finally, we close in Section \ref{sec:conclusion} with a summary of conclusions and further work.\\

\section{Eliciting Uncertainty from Decision Making}\label{sec:Methods}
We introduce a method that combines concepts from Indirect Interrogation methods as well as those that use historical data, by forming a prior distribution from an expert's past decision-making task. We are concerned with an undesirable future event $A$. The expert wishes to prevent $A$ from occurring and considers a (preventative) decision $Y$. Let $X$ be the information that is available to the decision maker at the time. The expert is interested in being able to quantify the prior probability on $A|X$, that is, what is the probability that $A$ will occur given the available information $X$. Using the expert's past decisions, the decision process $Y|X$ can be modelled. We conjecture that given $X$, the uncertainty in the outcome of $Y$ reflects the experts' uncertainty on whether $A$ would occur or not if no preventative measures were taken. Therefore, $A|X$ and $Y|X$ are intimately related. For simplicity, we assume that the event $A$ is binary (\emph{occurs or not}), so also is the preventative decision $Y$ (\emph{prevention put in place or not}). Let $Y|X \sim Bernoulli(p)$ and $A|X \sim Bernoulli(q).$ To be able to model the decision-making process $Y|X$ accurately, the process should be repetitive (carried out often) and its outcomes and the information used to make the decision should be available.\\

Let $Y_i$ denote the decision made at the $i^{th}$ instance (hereafter referred to as a \emph{case}) and $X_i$ be the information used by the decision maker to make that decision. Suppose that the data on $n$ cases is available so that we have $\mathbf{Y} = \{Y_1, Y_2, \ldots,Y_n\}$ and $\mathbf{X} = \{X_1,X_2,\ldots,X_n\}.$ Let $\boldsymbol{\theta}$ be model parameters that link the decisions $\mathbf{Y}$ to the available information $\mathbf{X}$ such that $\mathbf{Y}\sim f(\mathbf{Y}|\mathbf{X},\boldsymbol{\theta}).$ Given, $\mathbf{Y}$, $\mathbf{X}$ and a prior distribution $\pi(\boldsymbol{\theta}),$ we can find the posterior distribution $\pi(\boldsymbol{\theta}|\mathbf{Y},\mathbf{X}).$ Assuming information on a sufficient number $n$ of similar cases and an appropriate model $f$, it is reasonable to believe that using the information $X^*$ for the next case, we could accurately predict the decision $Y^*$ that the decision maker is likely to make using the posterior predictive distribution.
\begin{equation} \label{eq:Posterior Predictive Distribution}
    P(Y^*|X^*,\mathbf{Y},\mathbf{X}) = \int P(Y^*|X^*,\boldsymbol{\theta}) \pi(\boldsymbol{\theta}|\mathbf{Y},\mathbf{X}) \, d\boldsymbol{\theta}.
\end{equation}

Let $A_i$ be the undesirable consequence for the $i^{th}$ case, which may or may not materialize. The data on (some or all of) past $A_i$ may be available, but that is not considered here at this stage. Since $Y_i$ is the preventative decision to mitigate the risk of $A_i$, it is clear that $Y_i$ reflects the decision maker's prediction on $A_i$. That is, that a preventative decision was put in place implies that the decision maker believes that $A_i$ is likely to occur. Similarly, if the preventative measures were not put in place, this would reflect the decision maker's belief that $A_i$ is unlikely to occur. That is, 
\begin{equation} \label{eq:sameAY}
    A_i|X_i \approxtext{d} Y_i|X_i.
\end{equation}

\begin{figure}[h]
		\centering
		\includegraphics[width=.5\linewidth]{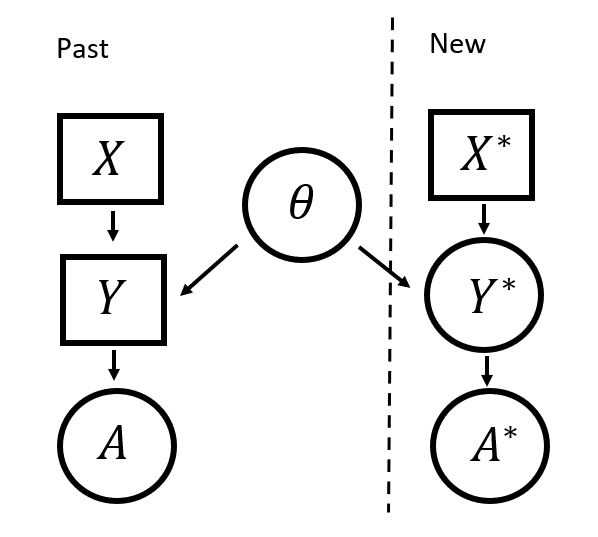}
		\caption{Influence Diagram for Eliciting Prior Distributions from Expert Decision Making}\label{figinfdiagram}
	\end{figure}

Therefore, given the information $X^*$ for the next case, the conditional predictive prior for $A^*$ can be approximated using the posterior predictive distribution in Equation \eqref{eq:Posterior Predictive Distribution}. That is,
\begin{equation} \label{eq:prior elicited}
\pi(A^*|X^*) \approx P(Y^*|X^*,\mathbf{Y},\mathbf{X}).
\end{equation}
See the accompanying influence diagram (Figure \ref{figinfdiagram}) that depicts the relationship between the variables.\\

%\vspace{5cm}

As an illustrative example, let $A$ be the event that a property in an industrial area will be burgled. This threat could be potentially mitigated by employing the services of a security consultant who would review the relevant information, $X$, make an assessment, $Y$, about the imminent risk and provide recommendations of security features that could be installed to prevent the threat from eventuating. If the data on $n$ recent property evaluations by the same consultant are available, then we can model the consultant's risk perception using a statistical model. The goal is to obtain the probability distribution of a new property being burgled using the relevant information available $X$. This probability distribution can be considered as an approximation to the consultant's prior probability distribution on whether the event $A$ will occur given $X$. \\

Note that our goal is not to accurately predict $A$. Instead, we want the model to accurately mimic the experts' decision-making process, and capture the experts' uncertainty about the event $A$, by considering the uncertainty in the model for the surrogate event $Y$. To ascertain whether the model is accurately mimicking the expert's decision-making process, an analyst can observe at least one of the measures of central tendency of the probability distribution of the parameter $p_i$, and assess whether it correctly predicts $Y_i$ in most of the cases (see Section \ref{sec:PM}). Moreover, we conjecture that the aleatory uncertainty captured by the model reflects the aleatory uncertainty of the expert on whether $A$ will occur or not given $X$. Our conjecture assumes that the decision maker recognizes that due to natural variability, an event may or may not occur even when it is very likely to occur and vice versa. \\

We will illustrate the use of this method with an example in Section \ref{sec:examples} using Bayesian logistic regression. Given $Y_i|X_i \sim Bernoulli(p_i)$, the logistic regression model, with a link function $g(.)$, is represented as, 
$$g(p_i) =  \theta_0 + \theta_1 x_1i + ...$$ \\
For example, with a simple logit link function and one predictor variable,
$$logit(p_i) = log(\frac{p_i}{1 - p_i}) = \theta_0 + \theta_1 x_i$$ 
\begin{gather}\label{eq:logitpeq}
    \Rightarrow p_i = \frac{exp(\theta_0 + \theta_1 x_i)}{1 + exp(\theta_0 + \theta_1 x_i)}
\end{gather}

A Bayesian approach is implemented by placing prior distributions on the model parameters, $\boldsymbol{\theta}= \{ \theta_0, \theta_1,...\}$. Sampling methods, such as MCMC methods, can be used to approximate the posterior distribution of $\boldsymbol{\theta}$. An analyst can select the prior distribution for model parameters and the sampling method and adjust them to build the most appropriate model (Section \ref{sec:PM}).  
To approximate the probability distribution for $p_i$ from this model, we can sample from the posteriors of the model parameters, $\pi(\boldsymbol{\theta} | \mathbf{Y}, \mathbf{X})$. These samples will be used in the model equation (for example Equation \ref{eq:logitpeq}) to obtain samples of $p_i$. An approach such as the methods of moments can then be used to fit a Beta distribution to these samples, which forms the elicited prior distribution of $q_i$ for the model $A_i|X_i \sim Bernoulli(q_i)$. \\   

There are many models that are used to predict rare or undesirable events, including Bayesian logistic regression models (e.g., for predicting recidivism \cite{tollenaar2013method,caulkins1996predicting,de2021modeling,schmidt1989predicting}). However, these models, to the best of our knowledge, have not yet been used to model expert decision-making or, to elicit an experts' prior distribution using the posterior predictive distribution. We reiterate that our goal is not to predict a rare or undesirable event, instead, we wish to capture the uncertainty surrounding said event occurring.\\

\section{Model Selection Diagnostics} \label{sec:PM}
To be able to elicit expert uncertainty accurately, we expect our model to behave like a decision-maker. We want it to be more uncertain when it sees data it has never seen before (wider distributions of $p_i$ that could be centered around 0.5) and less uncertain when it encounters familiar data (narrower distributions). 
Looking at the accuracy of the model is standard practice when assessing model performance (how accurately the model is predicting the response variable, $Y$, for a given test data set). If we wish to obtain the accuracy of a model which predicts the probability, $p_i$, of a binary decision, $Y_i$, labels are typically assigned as follows. If $p_i$ is less than 0.5 then the decision is labeled "no" and if $p_i$ is greater than 0.5 then the decision is labelled "yes" (or whatever the labels may be). When we are taking samples of $p_i$, it is common practice to take the mean of those samples as our estimate of $p_i$ that assigns labels. However, to assess how well the model captures the experts' thinking, model accuracy is not the only diagnostic that is of importance, as we must also take into consideration the variability of the elicited distributions and the uncertainty that they capture.\\
 \begin{figure}[h]
     \centering
     \begin{subfigure}[b]{0.3\textwidth}
         \centering
         \includegraphics[width=\textwidth,height=15cm,keepaspectratio]{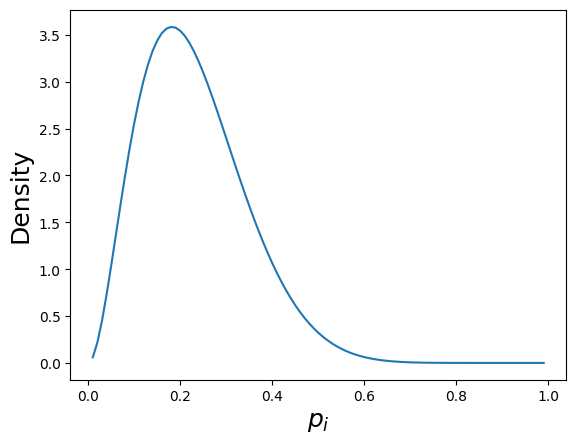}
         
         \label{fig:pm1}
     \end{subfigure}
     \hfill
     \begin{subfigure}[b]{0.3\textwidth}
         \centering
         \includegraphics[width=\textwidth,height=15cm,keepaspectratio]{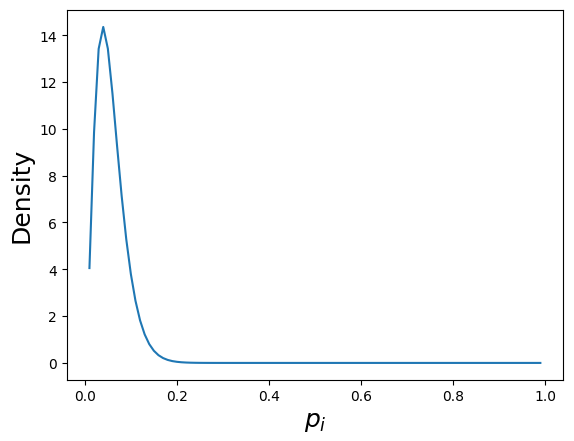}
         
         \label{fig:pm2}
     \end{subfigure}
     \hfill
     \begin{subfigure}[b]{0.3\textwidth}
         \centering
         \includegraphics[width=\textwidth,height=15cm,keepaspectratio]{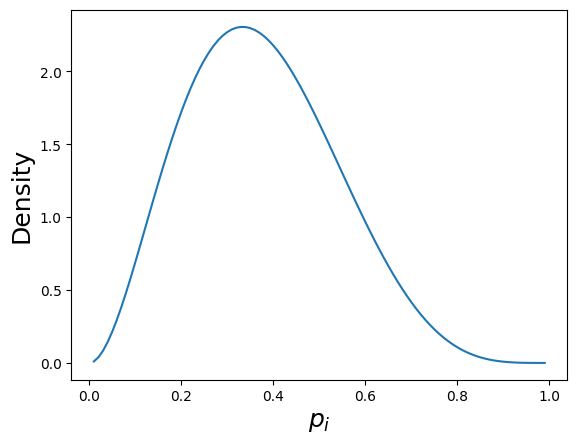}
         
         \label{fig:pm3}
     \end{subfigure}
     \centering
        \caption{Distributions of $p_i$ for three different individuals that would obtain the same label assigned based on mean probability prediction.}
        \label{fig:pmeg}
\end{figure}
It is easy to show that using the mean, of the sampled $p_i$ values, to assign labels for model accuracy may not give a fair representation of the variability of the distributions. For example, Figure \ref{fig:pmeg} shows three distributions where the model would assign the same label if the means of $p_i$ were used for assigning labels. However, we can see that using the mean does not accurately capture the difference in variability of the distributions and that using the median or the mode of the posterior predictive distribution would have assigned labels differently. We could also gain further insights by looking at the credible intervals of the distributions and assigning labels on whether the value of $p_i$ needed to assign a certain label, lies within the credible interval. The credible interval also allows an analyst to assess the uncertainty of the elicited distributions, which is of importance when selecting an appropriate model. If the credible interval is wide and contains 0.5, then we can assume that our expert is fairly uncertain, and if it is narrow and on either side of 0.5, we can assume they are fairly certain. In the same way, the area \emph{area under curve} (AUC) of the distribution can be used. To further assess the model's capabilities to capture uncertainty, an analyst can observe the entropy of the elicited distributions. If the entropy value is close to zero, then we assume the expert is fairly certain; if it is close to one, then we assume they are fairly uncertain. Assessing whether or not the model is behaving appropriately is case specific. If the analyst knows the decision making task has a lot of uncertainty, then they would expect high entropy values and will need to assess the trade-off between high entropy and high accuracy values. However, if the task is fairly certain, involving black and white responses, then we would expect low entropy values and aim for high accuracy from our model. \\

\begin{table}[hbt!]
	\caption{\label{TablePM} Model diagnostics we suggest to help select an appropriate model for prior elicitation.}.
	\centering
	\begin{tabular}{p{.3\textwidth}  p{.7\textwidth}}
 \hline
		\toprule
		%\multicolumn{2}{c}{Part}                   \\
		%\cmidrule(r){1-2}
		\textbf{Name}     & \textbf{Description}\\ \hline
		\midrule
		\emph{Mean Accuracy} & Percentage of correct predictions the model makes by using the mean of the sampled probabilities $p_i$.    \\
		\hline
 \emph{Mode Accuracy} & Percentage of correct predictions the model makes by using the mode of the sampled probabilities $p_i$.\\ \hline
 \emph{Median Accuracy} & Percentage of correct predictions the model makes by using the median of the sampled probabilities $p_i$.\\ \hline
 \emph{Area Under Curve (AUC) Accuracy} & Percentage of correct predictions the model makes by taking the largest area either side of 0.5 as the measure to form the model prediction. \\
 \hline
 \emph{95\% Credible Interval (CI) Accuracy} & Percentage of correct predictions the model makes by observing the 95\% CI of $p_i$. If the 95\% CI contains 0.5 then the assigned label can be either "Accept" or "Reject" and is a correct prediction. If the 95\% CI is contained below 0.5 and the true label is "Accept" then it is a correct prediction. If the 95\% CI is contained above 0.5 and the true label is "Reject" then it is a correct prediction.\\
 \hline
 \emph{Percentage of the 95\% CI correct predictions that contain 0.5.} & This will allow the analysts to see how many central distributions are elicited. \\
 \hline
 \emph{Percentage of the 95\% CI correct predictions that are either side of 0.5.} & This will allow the analysts to see how many skewed distributions are elicited. \\
 \hline
 \emph{F-Score \cite{sasaki2007truth}} & A measure which shows the specificity (true negative rate) and sensitivity (true positive rate) of the model. The mean of the samples of $p_i$ is used to assign labels. The highest possible value of an F-score is 1.0, indicating perfect specificity and sensitivity, and the lowest possible value is 0, if either the specificity or the sensitivity is zero. $$F = 2 . \frac{specificity . sensitivity}{ specificity + sensitivity}$$. \\
 \hline
 \emph{Confusion Matrix \cite{fawcett2006introduction}} & Shows the percentage of the mean predictions by whether the prediction is a true negative, true positive, false negative or false positive, showing the specificity and sensitivity of the model. The mean of the samples of $p_i$ is used to assign labels. \\
 \hline
 \emph{Entropy \cite{mackay2003information}} & A measure of the amount of uncertainty in a distribution. A narrow distribution will give a value close to zero (showing a certain prediction), and a wide distribution will give a value close to 1 (showing an uncertain distribution). To make sure the model is behaving correctly, it will be helpful to observe a histogram of all entropy values for the training set, as well as observing the histograms of the entropy values of correct and incorrect predictions separately. \\
 
 \hline
 \emph{Calibration Plot} & A calibration plot shows how well the prediction probabilities match the true percentage probabilities of the data. %A well calibrated model will show the blue line directly on the black line.
 The mean of the samples of $p_i$ is used as prediction probabilities. \\
 \hline
		\bottomrule
	\end{tabular}
\end{table}

These suggested diagnostics help an analyst assess the performance of the model, without looking at every single distribution produced. We advise analysts to look at multiple different model diagnostics to make sure the model is suitable for the task of prior elicitation and, also, to ensure they have a well-fitted model (Table \ref{TablePM}). The analyst's goal should be to maximise the model's accuracy (how well it is predicting the response for a given data set) while also producing distributions that accurately capture uncertainty.\\

\section{Example}\label{sec:examples}
Let $A$ be the event that a prisoner commits a crime upon release from prison. Information on a specific prisoner re-offending is limited and often censored, as we only know if a released prisoner commits a crime if they were caught. However, there exists an expert decision-making process that can be used to infer a prior distribution on the event $A$. This is the parole board hearing process. The parole board considers a report from a prisoner's case worker and decides whether or not to give a prisoner parole. When making a decision, the parole board is already taking into consideration the risk of the prisoner re-offending upon release, so this decision-making process can be used to infer a prior distribution on $A$. For example, if parole is not granted, this implies that the risk of re-committing a crime for an individual is high.\\

\subsection{Data}
We use a publicly available data set from the New York State Parole Board's interview calendar made available by The Parole Hearing Data Project \footnote{Data source \url{https://github.com/rcackerman/parole-hearing-data}}. This data set contains information on the prisoner, the hearing process, and the final decision\footnote{Data library \url{https://publicapps.doccs.ny.gov/ParoleBoardCalendar/About?form=datadef##Decision}}. It has 46 variables in total. We choose to take a subset of this data set by only looking at the initial parole board interviews. That is, the first time a prisoner appears before the parole board. The final data set has 9580 observations (Not Granted - 6962, Granted - 2618). The variables selected for our model are shown in Table \ref{tab::Variables}. Variables were selected based on their perceived relevance to the decision and if a variable had no impact on model performance it was removed. The posterior of each variable was also observed to see if the 95\% credible interval contained zero (meaning it has little to no impact on the model).\\

\begin{table}[hbt!]
	\caption{\label{tab::Variables}Variable names and descriptions}
	\centering
	\begin{tabular}{l  p{.7\textwidth}}
 \hline
		\toprule
		%\multicolumn{2}{c}{Part}                   \\
		%\cmidrule(r){1-2}
		\textbf{Variable Name}     & \textbf{Variable Description}\\ \hline
		\midrule
		\emph{Parole Board Decision} &  Simplified labels to a binary response: Granted = \{Open Date, Granted, Paroled\}, Not Granted = \{Denied, Not Granted\}.  \\
		\hline
 \emph{Gender} & Male, Female\\ \hline
 \emph{Ethnicity} & Black, White, Hispanic, Other \\ 
\hline
 \emph{Age} & Years from birth date to interview date. \\
 \hline
 \emph{Crime 1 Class} & Felony codes A, B, C, D and E. A felonies being the most serious and E felonies being the least serious.\\\hline
 \emph{Number of Years to Release Date} & Years from interview date to release date. \\\hline
 \emph{Number of Years to Parole Date} & Years from interview date to parole eligibility date.  \\\hline
 \emph{Aggregated Maximum Sentence} & Maximum aggregated amount of time a prisoner must serve for the crimes they are convicted of.\\\hline
 \emph{Aggregated Minimum Sentence} & Minimum aggregated amount of time a prisoner must serve for the crimes they are convicted of. \\\hline
 \emph{Crime Count} & Number of crimes a prisoner was convicted of under the given sentence (not all criminal history, just crimes for the current prison stay).\\\hline
 \emph{Crime 1 Conviction} & Simplified down to the following set: \{Possession: Crimes involving possession of an illegal substance or firearm; Grand Larceny: taking of goods in excess of \$1000; Assault: Crimes involving assault, excl. sexual assault; DWI: Driving under the influence of drugs or alcohol; Court: Crimes involving court proceedings(e.g., perjury, contempt); Sale: Crimes involving sale of an illegal substance or firearm; Sexual: Any sex related crime (e.g., sexual assault, rape); Fake: Crimes where an individual has faked something (e.g., forgery, identify theft);  Death: Any crime where an individual has caused death excl. murder (e.g., manslaughter, homicide); Stalking: including surveillance and harassment; Conspiracy, Murder, Robbery,  Arson, Fraud, Kidnapping, Other: All other crimes which do not come under any of the other labels\}. Reducing categories in this way is common practice in statistics and is  done throughout crime modelling \cite{ABS}.\\\hline %(e.g. \cite{dahal2020analysis,khan2022predicting})%
		\bottomrule
	\end{tabular}
	
\end{table}

\subsection{Model}
We wish to model the Parole Board Decision (response variable) using all other variables as explanatory variables (Table \ref{tab::Variables}). Numeric variables are standardised and categorical variables are changed to dummy variables.
The model is fitted and posterior distributions are found on a training data set that consists of 80\% of the full data set (7664 observations). The performance measures are assessed for a test data set of observations the model has never seen. The test data set consists of the remaining 20\% of the full data set (1916 observations). For a more accurate picture of how the model behaves, we randomly sampled five different testing and training sets and fitted the model separately in each case. We then took the average of the five different accuracy readings produced to get the final values.
The structure of the model is shown in Equation \ref{eq:modeleq}.\\

\begin{eqnarray}\label{eq:modeleq} 
Decision_i  & = & \beta_0 + \beta_1 \times gender\_male_i + \beta_2 \times age_i + \beta_3 \times num\_years\_release_i  + \beta_4 \times num\_years\_parole_i  \nonumber\\
 &  & + \  \beta_5 \times crime\_count_i  + \beta_6 \times agg\_min\_sent_i + \beta_7 \times  agg\_max\_sent_i + \beta_8 \times eth\_hispanic_i \nonumber\\
 &  & + \  \beta_9 \times eth\_white_i + \beta_{10} \times eth\_other_i + \beta_{11} \times crime\_class\_B_i + \beta_{12} \times crime\_class\_C_i \nonumber\\
 &  & + \  \beta_{13} \times crime\_class\_D_i + \beta_{14} \times crime\_class\_E_i + \beta_{15} \times crime\_conviction\_assault_i  \nonumber\\
 &  & + \  \beta_{16} \times crime\_conviction\_burglary_i + \ldots \\
  p_i & = & \frac{1}{1 + e^{Decision_i}} \nonumber
\end{eqnarray}

All parameters were initialised with a $Normal(0,0.001)$ prior. All trace plots of the parameters were acceptable.\\

\subsection{Model Diagnostics}
\begin{figure}[h]
     \centering
     \begin{subfigure}[b]{0.3\textwidth}
         \centering
         \includegraphics[width=\textwidth,height=15cm,keepaspectratio]{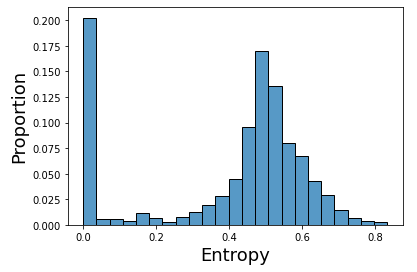}
         \caption{Histogram of the entropy for all test predictions.\\}
         \label{fig:entroeg1}
     \end{subfigure}
     \hfill
     \begin{subfigure}[b]{0.3\textwidth}
         \centering
         \includegraphics[width=\textwidth,height=15cm,keepaspectratio]{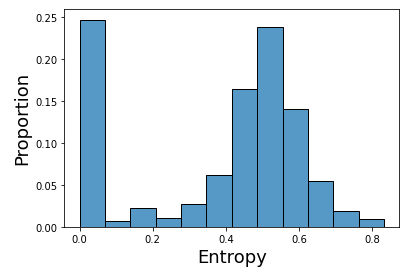}
         \caption{Histogram of the entropy for test predictions where the model made a correct prediction.}
         \label{fig:corentroeg1}
     \end{subfigure}
     \hfill
     \begin{subfigure}[b]{0.3\textwidth}
         \centering
         \includegraphics[width=\textwidth,height=15cm,keepaspectratio]{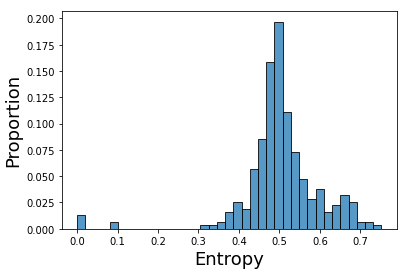}
         \caption{Histogram of the entropy for test predictions where the model made an incorrect prediction.}
         \label{fig:incorentroeg1}
     \end{subfigure}
        \caption{Entropy Plots}
        \label{fig:entroplotseg1}
\end{figure}

\begin{figure}[h]
     \centering
     \begin{subfigure}[b]{0.45\textwidth}
         \centering
         \includegraphics[width=\textwidth,height=10cm,keepaspectratio]{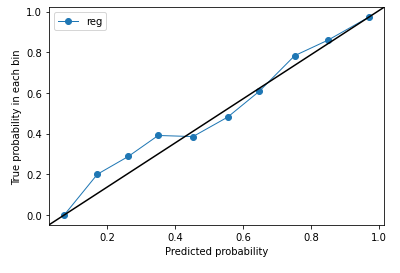}
         \caption{Calibration Plot}
         \label{fig:eg1cal}
     \end{subfigure}
     \hfill
     \begin{subfigure}[b]{0.45\textwidth}
         \centering
         \includegraphics[width=\textwidth,height=10cm,keepaspectratio]{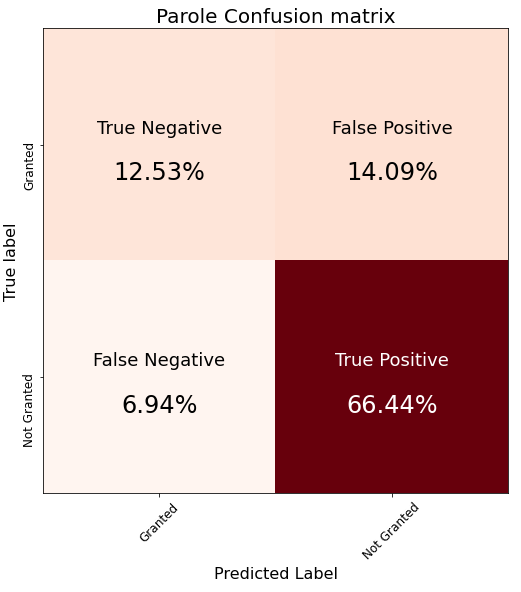}
         \caption{Confusion Matrix}
         \label{fig:eg1con}
     \end{subfigure}
     \hfill
    
        \caption{Other performance plots}
        \label{fig:eg1performanceplots}
\end{figure}

Accuracy readings were taken for the five different test sets and can be found in Table \ref{tab::e1accuracy}. The model obtains about 79\% classification accuracy overall. The CI accuracy is approximately 84\%, with 87\% of the CIs being on either side of 0.5, showing that the model is making more certain predictions than predictions that could be either "Granted" or "Not Granted" (corresponding to CI's containing 0.5). The F-score is around 0.87, which is close to one, showing that the model has relatively good specificity and sensitivity. Figure \ref{fig:entroeg1} shows the entropy of all observations in a single test set\footnote{NB:outputs were similar for all five test sets}. There are two peaks, one around zero and another around 0.5. From this, we can conclude that our model has some very certain predictions (peak around zero) and some less certain or very uncertain predictions (peak around 0.5). To gain further insight into the behaviour of our model in terms of entropy, Figure \ref{fig:corentroeg1} displays the entropy of correct predictions the model made and Figure \ref{fig:incorentroeg1} displays the entropy of incorrect predictions. We can see that for incorrect predictions the large peak at zero is not present (Figure \ref{fig:incorentroeg1}), whereas it is still present for correct predictions (Figure \ref{fig:corentroeg1}) meaning our model is less certain with its predictions when it is incorrect. The model looks relatively well-calibrated to the data (Figure \ref{fig:eg1cal}). The confusion matrix (Figure \ref{fig:eg1con}) shows that the model has a high true positive rate, that is, the model is predicting "Not Granted" well, which is to be expected due to the disproportionate amount of "Not Granted" versus "Granted" parole decisions in the data-set. Overall, we believe the model show acceptable behaviour for the proposed task.\\

\begin{table}[h] 
	\caption{\label{tab::e1accuracy} Average performance measures from five models.}
	\centering
	\begin{tabular}{l  p{.1\textwidth}}
 \hline
		\toprule
		%\multicolumn{2}{c}{Part}                   \\
		%\cmidrule(r){1-2}
		\textbf{Accuracy Measure}     & \textbf{Average} \\\hline
		\midrule
		\emph{Mean Accuracy} &  79.538\%  \\
		\hline
 \emph{Mode Accuracy} & 79.498\%\\ \hline
 \emph{Median Accuracy} & 79.51\% \\ 
\hline
 \emph{AUC Accuracy} & 79.488\% \\
 \hline
 \emph{95\% CI Accuracy} & 84.542\%\\\hline
 \emph{Percentage of the 95\% CI correct predictions that contain 0.5} & 12.832\% \\\hline
 \emph{Percentage of the 95\% CI correct predictions that are either side of 0.5} & 87.164\%  \\\hline
 \emph{F-Score} & 0.867 \\
 \hline
		\bottomrule
	\end{tabular}
	
\end{table}

\subsection{Elicited Prior Distribution}
After selecting the appropriate model, we can now obtain the elicited prior distribution for a new case. To produce a distribution of expert uncertainty for a single case, we obtain samples of $p_i$, the probability of a prisoner re-committing a crime, using the available information on the prisoner. We do this by sampling 100 times from the posterior distributions of the model parameters. These samples are then used to calculate samples of $p_i$, the probability of a decision $Y_i|X_i$. Then, the method of moments is used to fit a beta distribution to the samples of $p_i$, producing a final distribution capturing uncertainty. An analyst can also choose to fit other distributions to the data by MLE. They can then select the best distribution by the Kolmogorov-Smirnov test \cite{massey1951kolmogorov}.\\

\begin{figure}[h]
     \centering
     \begin{subfigure}[b]{0.3\textwidth}
         \centering
         \includegraphics[width=\textwidth,height=15cm,keepaspectratio]{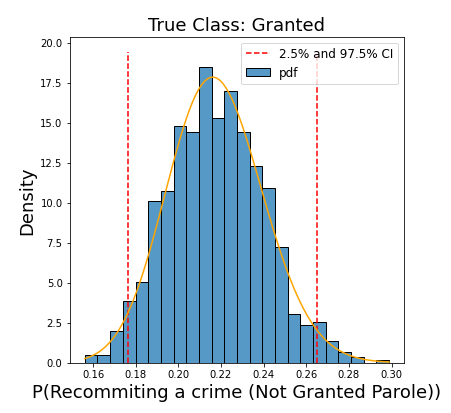}
         \caption{Prisoner 1}
         \label{fig:eg1in1}
     \end{subfigure}
     \hfill
     \begin{subfigure}[b]{0.3\textwidth}
         \centering
         \includegraphics[width=\textwidth,height=15cm,keepaspectratio]{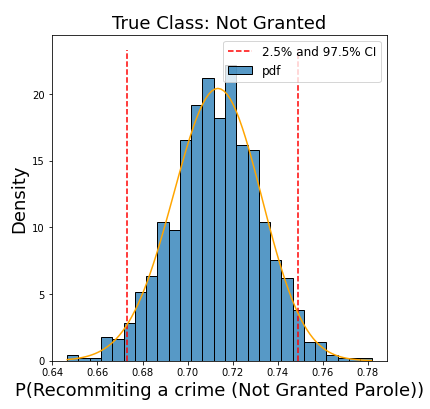}
         \caption{Prisoner 2}
         \label{fig:eg1in2}
     \end{subfigure}
     \hfill
     \begin{subfigure}[b]{0.3\textwidth}
         \centering
         \includegraphics[width=\textwidth,height=15cm,keepaspectratio]{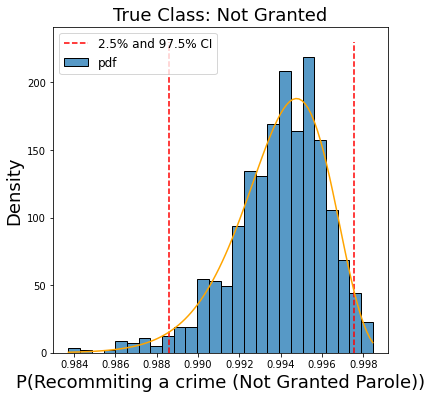}
         \caption{Prisoner 3}
         \label{fig:eg1in3}
     \end{subfigure}
        \caption{Prior distributions for three different prisoners}
        \label{fig:eg1inmatesoutput}
\end{figure}

Consider three prisoners: Prisoner 1, Prisoner 2 and Prisoner 3 (the prisoners' attributes are found in Table \ref{Tableeg1inmates}). The elicited prior distributions are shown in Figure \ref{fig:eg1inmatesoutput}. Prisoner 1 yielded a $Beta \sim (74.111, 266.202)$ prior distribution (Figure \ref{fig:eg1in1}). Prisoner 2 yielded a $Beta \sim (382.491, 154.224)$ prior distribution (Figure \ref{fig:eg1in2}). Prisoner 3 yielded a $Beta \sim (1181.395, 7.210)$ prior distribution (Figure \ref{fig:eg1in3}). These elicited distributions can now be used as prior distributions for recidivism for the given individuals and can be used to aid further decision-making.  \\

\begin{table}[h] 
	\caption{\label{Tableeg1inmates}: The prisoners' attributes used in Example 1}
	\centering
	\begin{tabular}{l  p{.1\textwidth} p{.1\textwidth} p{.1\textwidth}}
 \hline
		\toprule
		%\multicolumn{2}{c}{Part}                   \\
		%\cmidrule(r){1-2}
		\textbf{Attribute}     & \textbf{Prisoner 1} & \textbf{Prisoner 2} & \textbf{Prisoner 3}\\\hline
		\midrule
	
Age: & 34 years & 23 Years & 29 years\\ \hline
Number of years to release date: & 0 years & 0 years & 1 year\\ \hline
Number of years to parole date: & 0 years & 0 years & 0 years \\\hline
Aggregated Maximum Sentence: & 3 years & 3 years & 4 years \\\hline
Aggregated Minimum Sentence: & 1 year & 1 year & 1 years\\\hline
 Gender: & Male & Male & Male\\\hline
 Ethnicity: & White & Black & White\\ \hline
 Crime Count: & 1 & 1 & 2\\ \hline
 Crime 1 Conviction: & Burglary & Possession & DWI\\ \hline
 Crime 1 Class: & D & E & E\\ \hline
 Decision: & Granted & Not Granted & Not Granted\\ \hline
		\bottomrule
	\end{tabular}
	
\end{table}

\subsection{Influential Variables}
For this example, we have shown how an analyst can elicit a prior distribution from an expert decision-making process using tabular data. However, can an analyst trust that this elicited distribution is reliable? Can they trust the expert's decisions? Could some variables be wrongly influencing decisions? We chose to consider these questions by exploring variables seen in the decision-making process that should not have a cause-effect relationship with the decision. The variables we chose to explore are ethnicity, gender, and age. To explore the effect of these variables, we first created models without these variables and compared them to the original model. Each model was run five times with different testing and training data sets to produce an average of all accuracy measures. \\

\begin{figure}[h]
     \centering
     \begin{subfigure}[b]{0.45\textwidth}
         \centering
         \includegraphics[width=\textwidth,height=10cm,keepaspectratio]{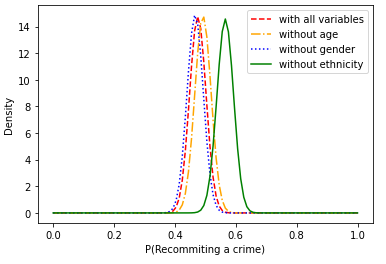}
         \caption{Prisoner 4}
         \label{fig:eg1modin1}
     \end{subfigure}
     \hfill
     \begin{subfigure}[b]{0.45\textwidth}
         \centering
         \includegraphics[width=\textwidth,height=10cm,keepaspectratio]{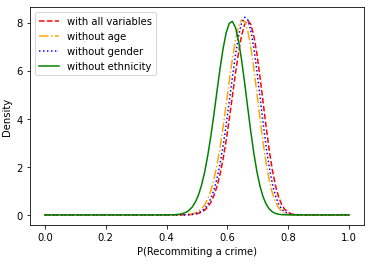}
         \caption{Prisoner 5}
         \label{fig:eg1modin2}
     \end{subfigure}
     \hfill
    
        \caption{Elicited distributions for the four different models for different prisoners.}
        \label{fig:eg1modelsinmatesoutput}
\end{figure}

\begin{figure}[h]
     \centering
     \begin{subfigure}[b]{0.45\textwidth}
         \centering
         \includegraphics[width=\textwidth,height=10cm,keepaspectratio]{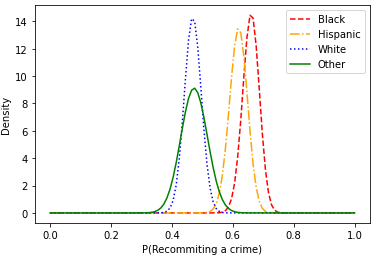}
         \caption{Prisoner 4}
         \label{fig:eg1ethin1}
     \end{subfigure}
     \hfill
     \begin{subfigure}[b]{0.45\textwidth}
         \centering
         \includegraphics[width=\textwidth,height=10cm,keepaspectratio]{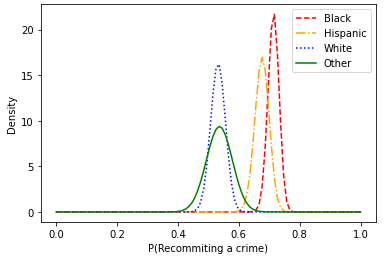}
         \caption{Prisoner 5}
         \label{fig:eg1ethin2}
     \end{subfigure}
     \hfill
    
        \caption{Elicited distributions for the same test point but with ethnicity variable changed}
        \label{fig:eg1eth}
\end{figure}

The model without ethnicity obtained the lowest average accuracy, and in fact, all five testing data sets gave lower accuracy than the full model (Table \ref{Tableeg1var}). It is also interesting to see that the model without Ethnicity has a higher percentage of 95\% CI correct predictions that contain 0.5. The model without age behaves roughly similar to the full model and the model without gender is only slightly less accurate. We also look at the behaviour of the elicited distribution of a test point from each model (Figure \ref{fig:eg1modelsinmatesoutput}). It can be seen that for each prisoner the full model and the models without age or gender perform similarly, however, the model without ethnicity produces a different distribution (Figure \ref{fig:eg1modelsinmatesoutput}). This finding is consistent for all prisoners considered.
We can further explore the impact of the variable ethnicity by using the full model and looking at a single prisoner and changing their ethnicity (Figure \ref{fig:eg1eth}). Again, there is a clear difference between the different ethnicity's elicited distributions. This shows us that ethnicity has an impact on the decision. Removing ethnicity from the model may reduce bias in the elicited prior distribution, but, it should be noted, that sometimes variables in tabular form can be representing other information that may be valuable to elicit an accurate prior distribution (confounding variables). For example, the variable ethnicity may be a proxy for socioeconomic status \cite{american2017ethnic}. This is a limitation of incomplete tabular data, as an analyst can only assume what this other information is. In this context, it is worth noting that there may be other methods that can go beyond tabular data and allow an analyst to use all the information a decision maker considered to elicit a prior distribution so that all the necessary information is kept in the model to elicit a prior distribution.\\

\begin{table}[h] 
	\caption{\label{Tableeg1var} Accuracy measures of models where the variables of interest are removed}
	\centering
 
	\begin{tabular}{p{.52\textwidth}  p{.12\textwidth} p{.12\textwidth} p{.12\textwidth} p{.12\textwidth}}
 \hline
		\toprule
		%\multicolumn{2}{c}{Part}                   \\
		%\cmidrule(r){1-2}
		\textbf{Accuracy \ Measure}    & \textbf{Full Model} & \textbf{Model without Ethnicity} & \textbf{Model without age} & \textbf{Model without gender}\\\hline
		\midrule
	
		\emph{Mean Accuracy} &  79.538\% & 78.286\%  & 79.77\% & 78.988\%\\
		\hline
 \emph{Mode Accuracy} & 79.498\% & 78.288\% & 79.488\% & 78.904\%\\ \hline
 \emph{Median Accuracy} & 79.51\% &  78.298\% & 79.72\% & 78.988\% \\ 
\hline
 \emph{AUC Accuracy} & 79.488\% & 78.298\% & 79.72\% & 78.978\% \\
 \hline
 \emph{95\% CI Accuracy} & 84.542\% & 85.564\% & 84.394\% & 84.3\% \\\hline
 \emph{Percentage of the 95\% CI correct predictions that contain 0.5} & 12.832\% &17.608\% & 12.074\% &  14.168\%\\\hline
 \emph{Percentage of the 95\% CI correct predictions that are either side of 0.5} & 87.164\% & 82.388\% & 87.904\% & 85.83\% \\\hline
 \emph{F-Score} & 0.867 & 0.857 & 0.868 & 0.86356 \\
 \hline
		\bottomrule
	\end{tabular}
	
\end{table}

\subsection{Summary}
This example shows how to elicit expert uncertainty present when considering whether a prisoner will re-commit a crime upon release, using a Bayesian logistic regression model to model parole board decision-making. The proposed process enables an analyst to also observe the impact of variables that may be influencing the decisions. The example has limitation; the parole board usually makes its decisions based on a report submitted by a prisoner's case worker. The only available data considered in the example was tabular data, which does not provide all the information that would be in the report. It would be interesting to see if modelling the report data would provide different results to those obtained above. Also, as with any elicitation method, there may be questions regarding the accuracy of the elicited prior distributions. The accuracy of elicited prior distributions is an ongoing concern of the prior elicitation field \cite{perala2020calibrating} and should be a continual path for future research.\\

\section{Conclusions and Future Work}\label{sec:conclusion} 
We introduce a new method to elicit prior distributions for an event, by modelling an expert decision-making task. We assume that a decision, $Y$, is closely related to the event $A$ so that samples from $P(Y|X,\theta)$, for different values of $\theta$, can be used to approximate the prior distribution for $A$ given a particular case $X$. This method allows an analyst to elicit a prior distribution from a real-world expert decision-making process, without the expert needing knowledge of probability concepts. This method can also be easily implemented for multiple experts where a decision is made in consensus because it models one decision, no matter if an individual or group makes the decision. We introduced this method with an example of recidivism using tabular data. This example used Bayesian logistic regression to model the parole board decision-making process. Once an appropriate model was fitted, samples from the posterior distributions of the parameters were taken to form a distribution that can be used as a prior distribution for recidivism. 
\\

Using this method also enables an analyst to explore variables that may be strongly influencing the decision-making process. What to do with this information should be a topic of future research. Should an analyst remove this information, or should it be shared with the experts to help train for future decision-making? A limitation of the logistic regression example considered in this paper is that the use of tabular data makes it challenging for an analyst to truly ascertain what is influencing the decision-making as this type of data only provides limited information and is often not what an expert would use to make their decisions. It would be interesting to explore modelling decision-making tasks that involve more complex data, such as images or reports. Basic statistical models cannot perform these tasks, instead, machine-learning approaches will have to be implemented.
Another limitation of the scenario considered in this paper is that we only consider decisions that have a binary outcome, however, there are many circumstances where decisions are not binary. There are ways to extend Bayesian logistic regression to the multinomial case, which should be explored further for prior elicitation. A concern in the field of prior elicitation is how accurate the elicited prior distribution is, further research could be taken to see how accurate this method of prior elicitation is and if there is a method to calibrate the elicited distribution (See example in \cite{perala2020calibrating}). If there exist cases where the outcome of the event $A$ has been observed, these could potentially be used to calibrate the elicited prior distribution. 
Overall, although we hope to have argued successfully that the proposed method is a promising candidate for prior elicitation in practical applications, further research should be performed to improve the practicality and generality of the approach.\\

\bibliographystyle{unsrt}
\bibliography{Main} 

\end{document}